# Spitzer Space Telescope Observations of the Nucleus of Comet 103P/Hartley 2

C.M. Lisse[1], Y.R. Fernandez[2], W.T. Reach[3], J.M. Bauer[3], M.F. A'Hearn[4], T.L. Farnham[4], O. Groussin[5], M.J Belton[6], K.J. Meech[7], C.D. Snodgrass[8]

*Under Revision for Re-Submission to PASP, 22-Jun-2009*

[1] Planetary Exploration Group, Space Department, Johns Hopkins University Applied Physics Laboratory, 11100 Johns Hopkins Rd, Laurel, MD 20723  carey.lisse@jhuapl.edu

[2] Department of Physics, University of Central Florida, 4000 Central Florida Boulevard, Orlando, FL. 32186-2385  yan@physics.ucf.edu

[3] Infrared Processing and Analysis Center, MS 220-6, California Institute of Technology, Pasadena, CA 91125  reach@ipac.caltech.edu, bauer@scn.jpl.nasa.gov

[4] University of Maryland, Astronomy Department, CSS 2341, College Park, MD 20742  ma@astro.umd.edu, farnham@astro.umd.edu

[5] Laboratoire d'Astrophysique de Marseille  olivier.groussin@oamp.fr

[6] Belton Space Exploration Initiatives, LLC, 430 Randolph Way, Tucson, AZ, 85716  mbelton@dakotacom.net

[7] Institute for Astronomy, University of Hawaii, 2680 Woodlawn Drive, Honolulu, HI 96822  meech@ifa.hawaii.edu

[8] European Southern Observatory, Alonso de Córdova 3107, Vitacura, Casilla 19001, Santiago de Chile, Chile  csnodgra@eso.org

# Abstract

We have used the Spitzer Space Telescope InfraRed Spectrograph (IRS) 22-μm peakup array to observe thermal emission from the nucleus and trail of comet 103P/Hartley 2, the target of NASA's Deep Impact Extended Investigation (DIXI). The comet was observed on UT 2008 August 12 and 13, while the comet was 5.5 AU from the Sun. We obtained two 200-frame sets of photometric imaging over a 2.7-hour period. To within the errors of the measurement, we find no detection of any temporal variation between the two images. The comet showed extended emission beyond a point source in the form of a faint trail directed along the comet's anti-velocity vector. After modeling and removing the trail emission, a NEATM model for the nuclear emission with beaming parameter of 0.95±0.20 indicates a small effective radius for the nucleus of 0.57 ± 0.08 km and low geometric albedo 0.028±0.009 (1σ). With this nucleus size and a water production rate of $3 \times 10^{28}$ molecules s$^{-1}$ at perihelion (A'Hearn *et al.* 1995) we estimate that ~100% of the surface area is actively emitting volatile material at perihelion. Reports of emission activity out to ~5 AU (Lowry *et al.* 2001, Snodgrass *et al.* 2008) support our finding of a highly active nuclear surface. Compared to Deep Impact's first target, comet 9P/Tempel 1, Hartley 2's nucleus is one-fifth as wide (and about one-hundredth the mass) while producing a similar amount of outgassing at perihelion with about 13 times the active surface fraction. Unlike Tempel 1, it should be highly susceptible to jet driven spin-up torques, and so could be rotating at a much higher frequency. Since the amplitude of non-gravitational forces are surprisingly similar for both comets, close to the ensemble average for ecliptic comets (Yeomans *et al.* 2004), we conclude that comet Hartley 2 must have a much more isotropic pattern of time averaged outgassing from its nuclear surface. Barring a catastrophic breakup or major fragmentation event, the comet should be able to survive up to another 100 apparitions (~700 yrs) at its current rate of mass loss.

## 1. Introduction

The Deep Impact (DI) mission, the eighth in NASA's Discovery Program, encountered comet 9P/Tempel 1 on UT 2005 July 04 (A'Hearn *et al.* 2005). Post encounter, the DI flyby spacecraft was re-targeted towards comet 103P/Hartley 2 via a close Earth flyby in December 2007. Now called the Deep Impact Extended Investigation (DIXI), the flyby spacecraft will encounter comet Hartley 2 on UT 2010 November 04, flying within 700 km of that comet's nucleus at a relative velocity of 12.3 km s$^{-1}$, 6 days after the comet's perihelion passage. Mission success depends critically on the ability of the DIXI spacecraft to navigate through the comet's extended atmosphere and image a kilometer-sized nucleus of unknown albedo, and planning for such a flyby requires knowing as much about the nucleus as possible well beforehand. Previous estimates of the nucleus's radius and albedo using mid-infrared imaging observations are uncertain by 40% at the 95% confidence limit (Groussin *et al.* 2004b) owing to interfering coma. Visible-wavelength observations of the comet when it was almost 5 AU from the Sun likewise showed some residual coma (Licandro *et al.* 2000, Lowry and Fitzsimmons 2001, Snodgrass *et al.* 2008). Here we present new Spitzer Space Telescope (Werner *et al.* 2004; hereafter *Spitzer*) observations of comet Hartley 2 taken when the comet was far from the Sun and the nuclear emission less affected by coma. These measurements take advantage of *Spitzer's* unprecedented sensitivity to perform a new characterization of the comet's nucleus.

## 2. Observations

**2.1. *Spitzer* Scheduling and Instruments.** We used the imaging mode of the InfraRed Spectrograph (IRS; Houck *et al.* 2004) aboard *Spitzer* for these observations despite its small (1.4'-by-0.9') field of view because (a) the ephemeris error was expected to be small (arcsecond scale), (b) that instrument was scheduled earlier in comet Hartley 2's observability window, when the comet was brighter due to smaller *Spitzer*-centric distance, and (c) the Multiband Imaging Photometer for the Spitzer (MIPS) instrument was scheduled for a time when the comet was near FY Lib, a 42-Jy MIII star that could have made background subtraction near the comet problematic. We only used the "red" (22 micron) peakup (PU) array on IRS since the comet was expected to be undetectably faint in the "blue" (16 micron) PU band. Observations started at UT 2008 August 12 23:56 and ended on August 13 02:30, when the comet was 5.457 AU from the Sun, 4.899 AU from *Spitzer*, and at a phase angle of 9.5º; this was 806 days before perihelion and 372 days after aphelion.

**2.2 Data Collection and Reduction.** The observations were performed as two separate pointings of the spacecraft, separated by about 75 minutes. At each pointing, we obtained 40 cycles of 5-point dithered exposures, with each exposure having 14 seconds of integration time. Each pointing's set of 200 images was processed by the Spitzer Science Center's pipeline version 18.1.0 to produce "basic calibrated data" (BCD) images. These images were flux-calibrated by the pipeline processing. We further processed these BCD images to remove instrumental background and artifacts. Then the 200 images in each set were registered in the comet's reference frame since the telescope was not tracking the comet during the observations. Smearing within each exposure was negligible. Finally, we stacked the frames to produce two final images that each have an effective exposure time 2800 seconds. The two images are shown in Figure 1.

## 3. Results and Discussion

**3.1 Photometry.** Both images contain a point source object that has an extended linear feature pointing to the west. The apparent sky-plane velocity of this object from one image to the next is consistent at the subpixel level with the motion of comet Hartley 2 as predicted by the ephemeris generated by the JPL Horizons service[1]. The linear feature lies at a position angle (PA 271º) that is consistent with the PA of the negative of the comet's projected heliocentric velocity; i.e., it lies along the comet's orbital path. The total extended source power in the trail was 25 - 30% of the total observed nuclear flux. No evidence for coma emission extending into an anti-solar comet tail at PA 110$^o$ was found.

---

[1] See URL http://ssd.jpl.nasa.gov/?horizons .

Finding the total photometric flux of Hartley 2's nucleus from our observations requires some attention to detail, since there is presumably emission from trail dust within the comet's head. This can be seen in the surface brightness profiles of the comet, shown in Fig. 2 as a function of cometocentric distance $\rho$. The pixels at the trail's azimuth are clearly seen above the background and in fact the trail surface brightness is comparable to that at the peak of the comet's head. To correct for the dust contribution and extract just the nucleus's flux, we employed an empirical fit to the images with a two-parameter, two-component morphological model array, M: $M = A_1 \times P + A_2 \times (D \otimes P)$. Here, P is the PSF array, D is an unconvolved dust model array, and $A_1$ and $A_2$ are scalar coefficients to be fit. The first term on the right-hand side represents the nucleus's contribution; the second, the dust's. The model was compared to an image array I over a fitting region and the goodness-of-fit was ascertained with the $\chi^2$ test, with $\chi^2 = \Sigma\ ((M_i - I_i)/\sigma)^2$ and the summation occurring over $i$, i.e. all pixels in the fitting region. The photometric error in a pixel is given by $\sigma$ and is 0.0156 MJy/sr in Fig. 1a and 0.0153 MJy/sr in Fig. 1b. The best-fitting coefficients then yielded the nucleus's flux $F = A_1 \times \int P\ d\Omega$, where $d\Omega$ is a pixel. A major advantage of this approach is that it removes any need to perform an uncertain aperture correction, which can become quite large for small apertures (factor of a few).

We obtained the PSF by using Tiny Tim/Spitzer[2], oversampling the PSF by a factor of 10 compared to actual PU pixels. To create the dust model array D, we decided to use the information in the data images themselves, where we see a trail that gets fainter approximately linearly as $\rho$ increases. To find the empirical slope of the trail surface brightness, we first determined the trail's azimuthal surface brightness profile within concentric, one-pixel wide annuli centered on the head's centroid. We then extracted pro-files in 19 annuli from ρ = 3 to 22 pixels in the first image (Fig. 1a), and in 10 annuli from ρ = 3 to 13 pixels in the second image (Fig. 1b). Each annulus's azimuthal profile was fit with a Gaussian to yield maximum surface brightness $S_{max}$. Fitting a line through the trend of $S_{max}$ with ρ produced a value for the slope $m$. (The resulting slopes from the two images were very similar, giving confidence in this method.) In the last step we produced model array D by creating a zero array (oversampled by a factor of 10), and then populating some of the pixels with trail flux. The trail in D is just 1-pixel wide, extends only on one side of the central pixel, and has a surface brightness that fades from the central pixel with slope equal to $m$.

---

[2] See URL http://ssc.spitzer.caltech.edu/archanaly/contributed/stinytim.

We executed the modeling by trying out different values of $A_1$ and $A_2$. Our model was able to fit both images more than adequately. The $\chi^2$ were acceptable even when we used a variety of fitting regions. The fitting regions included a circular aperture centered on the head's centroid of 2 to 4 PU pixels radius, plus a 20 to 40 degree-wide wedge that extended out another 4 to 8 PU pixels beyond that to cover the trail. For these regions, the reduced $\chi^2$ were typically 0.9 for Fig. 1a and 0.8 for Fig. 1b (with 20 to 80 degrees of freedom).

Using the best-fitting coefficients $A_1$ for various fitting regions, we find that the nucleus's flux in each image is the same, 0.100±0.018 mJy (error is 1σ). To within the errors of the measurement, we find no detection of any rotational modulation between the two images. We then applied a color correction of 1.023±0.002, derived using the appropriate expected color temperature of the nucleus (see below). Thus our final photometry for the total thermal emission from comet Hartley 2's nucleus in both images is 0.102±0.018 mJy.

It is important to note that this result depends crucially on array D. The assumption of having a trail that is just one oversampled pixel wide is justified, since the trail is made up of large, slow-moving grains that are localized in the orbit plane. At the time of observation, Spitzer was just 2.7º above comet Hartley 2's orbit plane (as seen from the comet), and furthermore one PU pixel subtended a full 6500 km (so an oversampled pixel was 650 km). More critical however is the assumption of the trail slope. Our formulation for D assumes that the trail maintains that surface brightness slope effectively all the way to the surface of the nucleus. This is by no means guaranteed, and we did try to vary the values of D's central pixels to gauge its effect on the resulting nuclear flux. By changing the slope within the central seeing disk, or by removing dust from some pixels, we were able to find models that adequately fit the images using nuclei as bright as 0.15 mJy. We were unable to extract a nucleus brighter than this and still retain a good fit, although the number of trail variations one can try is certainly large and we by no means sampled all possible choices. Nonetheless we consider this flux density to be an upper limit.

**3.2 Thermal Models and Nucleus Size.** The conversion from radiometry to physical properties – in particular effective radius, $R_{nuc}$ -- requires a thermal model, as discussed for asteroids by Lebofsky and Spencer (1989), Spencer *et al.* (1989), Harris and Lagerros (2002), and for comets by Lisse *et al.* (1999, 2005); Groussin and Lamy (2003); and Lamy *et al.* (2004). In particular, the thermal regime must be estimated so that an appropriate thermal model can be employed. The extreme cases are "slow" and "fast" rotators; a slow-rotator has either no rotation or a very short thermal energy emission time, so that every surface element is in instantaneous thermal equilibrium with the insolation. A fast rotator has a very short rotation period and/or very long thermal energy emission time, so that a surface element has an averaged temperature regardless of the instantaneous solar zenith angle. The thermal regime can be characterized by the parameter $\Theta$, defined by Spencer *et al.* (1989) as $\Theta = I \omega^{1/2} / \varepsilon \sigma T_{ss}^3$, where I is the thermal inertia in $J\ K^{-1}\ m^{-2}\ s^{-1/2}$, $\omega$ is the angular spin rate in rad sec$^{-1}$, $\varepsilon$ is the emissivity (assumed to be 0.95), $\sigma$ is the Stefan-Boltzmann constant, and $T_{ss}$ is the subsolar temperature in °K. Thermal inertia – equal to the geometric mean of thermal conductivity and volumetric heat capacity – is unknown for comet Hartley 2 but recent work on other comets and primitive icy bodies (Julian *et al.* 2000; Fernandez *et al.* 2002; Groussin and Lamy 2003, Lamy *et al.* 2004, Groussin *et al.* 2004a, 2007; Lisse *et al.* 2005) suggests that it is likely between 0 and 30 $J/K/m^2/s^{1/2}$. (For comparison, lunar regolith has I = 50 $J/K/m^2/s^{1/2}$ and solid rock has roughly 2500 $J/K/m^2/s^{1/2}$.) Here we adopt I=15 $J/K/m^2/s^{1/2}$ for comet

Hartley 2. The subsolar temperature goes as the inverse square root of heliocentric distance; a strongly-emissive, low-albedo plate at the subsolar point of a body at 1 AU will reach about 390 K; we adopt 390 K/√5.457 = 166 K for comet Hartley 2. The comet's rotation period is unknown, and our observations did not reveal any significant rotational modulation. If we assume a long rotation period (such as was seen for comet Tempel 1) of 41 hours (Lisse *et al.* 2005, A'Hearn et al 2005, Belton and Drahus 2007) then we find that $\Theta$ = 0.39, barely in the slow-rotator regime. If we assume a short rotation period of 6 hours (Toth and Lisse 2006) then $\Theta$ = 1.34, near the turnover point in thermophysical behavior between slow and fast rotators. This suggests that we must carefully account for heat conduction explicitly in a thermal model; neither extreme thermal scenario will suffice.

Our approach is then similar to that taken by Groussin *et al.* (2004a) in their study of Centaur 95P/Chiron. This model uses aspects of the Near-Earth Asteroid Thermal Model (NEATM; Harris 1998) but also accounts for some of the energy going toward volatile sublimation. For the Spitzer observations of comet Hartley 2, there is no detectable coma and thus we set the sublimation rate to zero. We choose a beaming parameter $\eta$ of 0.95; for comparison the ensemble average over about 50 nuclei is approximately 0.94 (Fernandez *et al.* 2008). We also choose a Bond albedo of 0.01. The result of our thermal modeling is that we find that the nucleus of comet Hartley 2 has an effective radius of 0.57±0.05 km (1$\sigma$). By including a wider range of possible values for $\eta$ -- say 0.75 to 1.15 -- we can estimate some of the systematic error due to model assumptions. We find that this boosts the 1$\sigma$ error bar somewhat and our final size estimate for comet Hartley 2 is $R_{nuc}$ = 0.57±0.08 km.

An effective radius of 0.57 km for comet Hartley 2 places it at the small end of the ecliptic comet (Levison 1996) nuclear size distribution (Fernandez *et al.* 2008) but close to the modal size for these comets. It is, however, about one-fifth the radius and thus roughly one-hundredth the mass of the DI spacecraft's first target, comet Tempel 1, which has a nuclear radius of 3.0 km (A'Hearn *et al.* 2005). It is possible, given that comet Hartley 2 is a small, normal carbon-chain abundance comet (unlike the typical

JFC), and that it only recently started orbiting to within 3 AU of the Sun (A'Hearn *et al.* 1995), that it represents an individual primordial cometesimal more than the compound, layered, geologically complex comet Tempel 1. Support for this idea comes from the TALPS model of cometary nuclei (Belton *et al.* 2007), which suggests that the Tempel 1 nucleus was created by the aggregation of many smaller individual proto-cometary planetesimals, Further support comes from an examination of the exposed layer deposits on the Tempel 1 surface (Thomas *et al.* 2007). We find that the equivalent radius of a cometary body that would cover half of Tempel 1's surface with a 20-meter thick layer in an accretive impact would be > 0.65 km (where the lower limit is achieved for a lossless impact with no compression). This size is very similar to the size found here for the Hartley 2 nucleus.

**3.3 Geometric Albedo Estimate.** We have formed an average geometric albedo estimate for the Hartley2 nucleus using our Spitzer derived nucleus size coupled with ESO/VLT FORS2 camera R-band observations of the comet on 28-Jul-2008 (Snodgrass *et al.* 2009). A value of $R(r_h = 1$ AU, $\Delta = 1$ AU, phase $= 0°)$ equal to $18.9 \pm 0.20$ mag was obtained in a somewhat crowded starfield for the nucleus' total scattered flux, assuming a phase coefficient $\beta = 0.035$ mag/deg. Using $p_v = (1.496 \times 10^8$ km$/ R_{nuc}(km))^2 * 10^{-0.4*(Rcomet(1,1,0)-RSun(0,1,0))}$ (Lisse *et al.* 1999), the Spitzer value for the effective radius from this work of $R_{nuc} = 0.57$ +/- 0.08 km, and the apparent R-band brightness of the Sun from the Earth $R_{Sun} = -27.1$, we find $p_v = 0.028 \pm 0.009$ (1$\sigma$).

The derived albedo is low for a solar system body, but consistent with the canonical $p_v = 0.04$ found for the ensemble average albedo of comets (Fernandez *et al.* 2001, Lamy *et al.* 2004). The error bars on the albedo estimate include the statistical uncertainties in R(1,1,0) and $R_{nuc}$, but not the systematic error due to the unknown rotational context of the VLT and Spitzer observations, nor any systematics due to the optical background subtraction. The former systematic is of uncertain effect, but the latter most likely has caused an over-subtraction of the background and an underestimation of the total scattered nuclear flux and albedo.

**3.4 Trail.** Our Spitzer observations of comet 103P/Hartley 2 demonstrate that the comet supports a trail structure (Figure 1). The trail surface brightness (Figure 2) of 0.036 ± 0.013 MJy/sr, coupled with a temperature at 5.5 AU for large, rapidly-rotating grains of 120 K, implies an optical depth for the trail $\tau \sim 2 \times 10^{-9}$, within a factor of 2 of the median for ecliptic comets (Reach *et al.* 2007). The apparent difficulty in detecting the trail in previous observations had not been due to the trail brightness, but to the bright and overlapping coma and tail resulting from activity of the nucleus while even 5 AU from the Sun. The presence of a ~20% silicate emission feature (Crovisier *et al.* 2000) also indicates that the dust emitted when the comet is active has a large proportion of grain surface area in micron-sized particles, which can easily obscure the larger trail particles.

Comparing the Spitzer image (Figure 1) to numerical dynamical simulations of the trail using the model described by Reach *et al.* (2007), we find that the trail length (extending beyond the edge of the image) and width match those predicted for mm-sized particles produced during the 2004 May perihelion passage. The width of the trail perpendicular to the orbit plane is $\sim 4 \times 10^4$ km (Figure 2), which when combined with the optical depth yields a mass density of mm-sized debris of $7 \times 10^{-20}$ g cm$^{-3}$. For a spacecraft flying through the trail perpendicular to the orbital arc, the expected number of impacts per square meter of spacecraft is of order $10^{-3}$. Thus the main dust hazard to the DIXI spacecraft in 2010 should be near-nuclear coma dust, rather than dust in the debris trail. Tempel 1's debris trail has a similar optical depth (Sykes & Walker 1992; Reach *et al.* 2007), so we predict that the spacecraft will encounter a similar large-particle environment as was encountered at Tempel 1, where four large particles (of μg to mg size) were encountered within 200 km of the nucleus (Lisse *et al.* 2006, A'Hearn *et al.* 2008). Current mission plans have the DIXI spacecraft approach to within 700 km of Hartley 2 in November 2010.

**3.5 Activity, Active Fraction, and Lifetime.** Comparing the behavior of the two Deep Impact mission targets is a useful exercise. While Hartley 2's nucleus is smaller – 0.57 km vs. 3.0 km (Thomas *et al.* 2007) – implying $(3.0/0.57)^2 = 27$ times less overall surface area, the two comets have similar gas production rates at perihelion. Comet Hartley 2 receives about twice the solar flux at perihelion since it approaches to 1.06 AU vs. 1.50 AU, but this still suggests comet Hartley 2 could have an active fraction, $x$, that is roughly 13 times larger than that of comet Tempel 1. Using the value of $x = 0.09$ for Tempel 1 from A'Hearn *et al.* (1995), this implies $x \sim 1.17$ for comet Hartley 2, or a fully active and emissive surface, since $0 < x < 1.0$ by definition. A more detailed calculation, based on our thermal modeling, yields a similar answer. A'Hearn *et al.* (1995) estimated the perihelion $Q_{H2O}$ to be $3 \times 10^{28}$ mol sec$^{-1}$. Adopting this value and using a temperature distribution with $I = 15$ J/K/m$^2$/s$^{1/2}$, we find that $x \geq 99\%$ at perihelion. The high active fraction estimates derived here are in agreement with the high end of the range given by Groussin *et al.* (2004b), who reported $0.3 < x < 1.0$.

Figure 3 compares the nuclear surface active area for comet Hartley 2 derived from this work to that found for the targets of the 85 comet photometric survey of A'Hearn *et al.* (1995). While the absolute value of the active area is comparable to that of other, relatively-small, kilometer-scale comets, such as 46P/Wirtanen (Farnham and Schleicher 1998, Groussin and Lamy 2003), the relative fraction of the nucleus surface that is actively sublimating is remarkably large compared to the ensemble average. It is possible that comet Hartley 2 is small enough that solar insolation can drive devolatilization from a good fraction of its remaining volume, explaining its high activity. Another possibility that must be considered is the existence of significant population of icy dust in the coma emitting water gas in an extended source. While ISO observations of the coma dust near perihelion (Crovisier *et al.* 2000) do not show evidence for cold icy grains, Lisse *et al.* (1999) showed for the case of C/1996 B2 (Hyakutake) that it takes only a small fraction of the total coma dust for a moderately active comet, on the order of a few percent, to provide orders of magnitude more surface area than the nuclear surface supporting water gas sublimation. Further, they found that a few percent by surface area of cold (150 – 200°K) icy dust would be very difficult to detect in contrast with the surrounding warm (300- 400°K) refractory dust in the coma.

Assuming a bulk density close to that found for comet Tempel 1, i.e., ~400 kg m$^{-3}$ (Richardson *et al.* 2007), a spherical comet, and the effective radius $R_{nuc} = 0.57$ km from this work, we find a total nuclear mass of ~3 x $10^{11}$ kg. Despite the apparently very high nuclear surface activity, comet Hartley 2 should be able to survive another ~$10^2$ apparitions at its current rate of coma emission and rate of total mass loss, ~$10^9$ kg/orbit (Lisse 2002). If, however, we allow for the possibility of future fragmentation and rotational breakup, as seen recently for the small prime nucleus of comet 73P/Schwassmann-Wachmann 3 ($R_{nuc}$ ~ 0.7 km; Toth *et al.* 2008, Weaver *et al.* 2008, Reach *et al.* 2009), then the mass loss can be greatly accelerated from its current rate. We thus consider ~$10^2$ apparitions, or ~700 yrs, to be an upper limit for the lifetime of comet Hartley 2 as a comet.

**3.6 Dynamical Considerations.** The relatively small radius and high water production rate of comet Hartley 2 suggest that its orbital and rotational dynamics may be quite distinct from a body with which we have more experience, like comet Tempel 1. In order to be specific we explore this idea assuming that the current spin period, $P_{spin}$, of comet Hartley 2 is near the median for Jupiter Family Comets (JFCs), ~0.5 day (Lamy *et al.* 2004). Given that the size of comet Hartley 2 is close to the median size of the JFC comets (Fernandez *et al.* 2008), we believe this to be a reasonable assumption.

The non-gravitational force vector, **A,** is a measure of the acceleration of the nucleus relative to the sun, *i.e*. body force/unit mass (Yeomans *et al.* 2004), due to the outflow of gas and dust. Since we expect water to dominate, we might also expect the magnitude of **A,** |**A**|, to scale as ~$Q_{H2O}/R_{nuc}^3$, where $Q_{H2O}$ is evaluated at its peak value. For comet Tempel 1, the peak $Q_{H2O}$ is 6 – 7 x $10^{27}$ mol sec$^{-1}$ (Schleicher 2007) and $R_{nuc}$ = 3.0 km (Thomas *et al.* 2007), while for comet Hartley 2 the peak $Q_{H2O}$ is 3 x $10^{28}$ mol.s$^{-1}$ (A'Hearn *et al.* 1995) and $R_{nuc}$ = 0.57 km (this work). Thus we might expect |**A**| to be about 700 times greater in comet Hartley 2 than comet Tempel 1. In fact, it is only 2.7 times larger, and both values are near the median for the JFC comets. (Current estimates of **A** can be found on the JPL Horizons website.) This suggests that either the non-gravitational forces acting on comet Tempel 1 were unusually high for such a large nucleus, or that the outgassing from the Hartley 2 surface occurs in a relatively isotropic spatial pattern, averaging out the reactive jet forces of material emitted from the nucleus.

Gas and dust outflows can also apply torques to the nucleus causing the spin of the nucleus to evolve. The e-folding time T for a change in the angular momentum (Jewitt 1997) goes as $\sim R_{nuc}^4/(P_{spin}*Q_{H2O})$. If Hartley 2's P = 0.5 day, then T ~ 0.2 yr; the very small moment of inertia of comet Hartley 2 could make it highly susceptible to torques on orbital timescales. For Tempel 1, with $P_{spin}$ = 1.7 day, T ~ 88 yr, much longer than the orbital period of 5.5 yrs and more than two orders of magnitude longer than the e-folding time for Hartley 2. A measure of the magnitude of the net torque being applied to the nucleus of comet Tempel 1 can be derived from the measured angular acceleration of the spin, 0.021 deg day$^{-2}$ (Belton and Drahus 2007); the net torque is roughly 1 x 10$^7$ kg m$^2$ s$^{-2}$ (Belton and Drahaus 2007, Belton *et al.* 2009). Since torque scales as ~ $R_{nuc}*Q_{H2O}$, this suggests that Hartley 2 would feel a comparable torque. If so, and given that the moment of inertia scales as $R_{nuc}^5$, comet Hartley 2 could be experiencing angular accelerations roughly 3400 times greater -- i.e. about 60 deg day$^{-2}$. Furthermore, with torques as high as noted above, forced precession rates as high as 0.9$^o$ day$^{-1}$ could be possible. Thus not only might make the spin state undergo substantial changes in the instantaneous spin rate, but also the direction of the spin pole might change by as much as ~30$^o$ in a single perihelion passage.

It is important to note that all this analysis also depends on the magnitude of the two nuclei's dimensionless moment arms, $k_T$ (Jewitt 1997), i.e. the extent to which outgassing from either nucleus is tangential vs. radial. The value of $k_T$ has not been estimated for many comets and is likely uncertain at the order-of-magnitude level. The spin changes hypothesized here for Hartley 2 would be mitigated if its $k_T$ is significantly lower than Tempel 1's.

## 4. Conclusions

From our Spitzer observations, all indications are that Deep Impact extended mission target 103P/Hartley 2 is a very small comet of effective radius 0.57±0.08 (1σ) km and typically low cometary geometric albedo of 0.028±0.009 (1σ). It is likely that it represents a primordial cometesimal more than the compound, layered, geologically complex comet 9P/Tempel 1, target of the Deep Impact prime mission. It exhibits an unusually high emission activity from a remarkably large proportion of its surface out to large heliocentric distances (~ 5 AU). It is possible that comet Hartley 2 is small enough that solar insolation can drive devolatilization from a good fraction of its remaining volume, explaining its high activity. While we found no discernable variation over the 2.7 hrs of our Spitzer observations in the emitted thermal flux of the nucleus due to rotational variation, with it small mass, moment of inertia, and high outgassing rate, comet Hartley 2 should be easily susceptible to changes in its rotation state. At its current total mass of $\sim 10^{12}$ kg and mass loss rate $-dM/dt = 10^9$ kg/orbit, comet Hartley 2 should be able to survive up to another $10^2$ apparitions or 700 yrs, assuming it continues its current rate of mass loss and suffers no catastrophic break up due to the generation of an excessive spin rate during this time.

## 5. Acknowledgements

This work is based on observations taken with the Spitzer Space Telescope, which is operated by JPL/Caltech under a contract with NASA. Support for this work was provided by NASA through an award issued by JPL/Caltech. This research made use of Tiny Tim/Spitzer, developed by J.E. Krist for the Spitzer Space Center. The authors would like to thank D.K. Yeomans for valuable discussions concerning the effects of non-gravitational forces on comets, and H.A. Weaver, for results from the latest optical observations of comet Hartley 2.

# 6. References

A'Hearn, M.F., Millis, R.L., Schleicher, D.G., Osip, D.J., Birch, P.V., 1995. The ensemble properties of comets: Results from narrowband photometry of 85 comets 1976-1992. *Icarus* **118**, 223-270.

A'Hearn, M.F., *et al.* 2005. Deep Impact: Excavating comet Tempel 1. *Science* **310**, 258-264.

A'Hearn, M.F., Belton, M.J.S., Farnham , T.L., Feaga, L.M., Groussin, O., Lisse, C.M., Meech, K.J., Schultz, P.H., and Sunshine, J.M. 2008. "Deep Impact and Sample Return", *Earth Planets Space* **60**, 61-66.

Belton, M.J.S., *et al.* 2007. The Internal Structure of Jupiter Family Cometary Nuclei from Deep Impact Observations: The Talps or Layered Pile Model. *Icarus* **187,** 332 - 344.

Belton, M.J.S., and Drahus, M., 2007. The accelerating spin of 9P/Tempel 1. *Bull.A.A.S.* **39**, 498.

Belton, M.J.S., and 33 colleagues 2009. Deep Impact, Stardust-NExT and the accelerating spin of 9P/Tempel 1. *Icarus*, submitted.

Crovisier, J., Brooke, T.Y., Leech, K., Bockelée-Morvan, D., Lellouch, E., Hanner, M.S., Altieri, B., Keller, H.U., Lim, T., Encrenas, S., Griffin, M., deGraauw, T., van Dishoeck, E., Knacke, R.F. 2000. The thermal infrared spectra of comets Hale-Bopp and 103P/Hartley 2 observed with the Infrared Space Observatory. In: Thermal Emission Spectroscopy and Analysis of Dust, Disks, and Regoliths, Eds. M.L. Sitko, A.L. Sprague, and D.K. Lynch, *Astronomical Society Pacific Conference Series* **196**, pp. 109-117.

Farnham, T. L. and D. G. Schleicher 1998. Narrowband photometric results for comet 46P/Wirtanen. *Astron. Astrophys*. **335**, L50-L55.

Fernandez, Y.R., Jewiit, D.C., and Sheppard, S.S., 2001. Low albedos among extinct comet candidates. *Astrophys. J.* **553**, L197–L200

Fernández, Y.R., Jewitt, D.C., and Sheppard, S.S. 2002. Thermal properties of centaurs Asbolus and Chiron. *Astron. J.* **123**, 1050-1055.

Fernandez, Y.R., Kelley, M., Lamy, P., Reach, W., Toth, I., Groussin, O., Lisse, C., A'Hearn, M., Bauer, J., Campins, H., Fitzsimmons, A., Licandro, J., Lowry, S., Meech, K., Pittichova, J., Snodgrass, C., Weater, H., 2008. Results from SEPPCoN, a survey to study the physical properties of the nuclei and dust of Jupiter-Family comets." *American Geophysical Union Spring Meeting 2008*, abstract #**P41A-08**.

Groussin, O. and Lamy, P. 2003. Activity on the surface of the nucleus of comet 46P/Wirtanen. *Astron. Astrophys.* **412**, 879 – 891.

Groussin, O., Lamy, P., and Jorda, L. 2004a. Properties of the Nuclei of Centaurs Chiron and Chariklo. *Astron. Astrophys.* **413**, 1163 – 1175.

Groussin, O., Lamy, P., Jorda, L., and Toth, I. 2004b. The nuclei of comets 126P/IRAS and 103P/Hartley 2. *Astron. Astrophys.* **419**, 375-383,

Groussin, O., A'Hearn, M.F., Li, J-Y, Thomas, P.C., Sunshine, J.M., Lisse, C.M., Meech, K.J., Farnham, T.L., Feaga, L.M., and Delamere, W.A. 2007. Surface temperature of the nucleus of comet 9P/Tempel 1. *Icarus* **187**, 16 - 25.

Harris, A. 1998. A thermal model for near-Earth asteroids. *Icarus* **131**, 291-301.

Harris A.W., and Lagerros, J.S.V. 2002. Asteroids in the thermal infrared. In *Asteroids III*. Bottke, W.F.,

Cellino, A., Paolicchi, P., Binzel, R.P. (Eds.). Univ. of Arizona Press, Tucson, p. 205 - 218.

Houck, J.R. *et al.* 2004. The infrared spectrograph (IRS) on the Spitzer Space Telescope. *Astrophys. J. Supp.* **154**, 18 - 24.

Jewitt, D. 1997. Cometary Rotation: An Overview. *Earth, Moon and Planets* **79**, 35 – 53.

Julian, W.H., Samarasinha, N.H., Belton, M.J.S., 2000. Thermal structure of cometary active regions: comet 1P/Halley. *Icarus* **144**, 160 – 171.

Lamy, P., Biesecker, D.A., and Groussin, O. 2003. SOHO/LASCO observation of an outburst of comet 2P/Encke at its 2000 perihelion passage". *Icarus* **163**, 142 – 149.

Lamy, P. *et al.* 2004 The sizes, shapes, albedos, and colors of cometary nuclei. In *Comets II*. Eds. M.C. Festou, H.U. Keller, and H.A. Weaver. University of Arizona Press, Tucson. pp. 223-264.

Lebofsky, L.A. and Spencer, J.R. 1989. Radiometry and thermal modeling of asteroids. In *Asteroids II*. Binzel, R.P., Gehrels, T., Matthews, M.S. (Eds). Univ. of Arizona Press, Tucson, AZ, 128 – 147.

Levison, H.F., 1996. Comet Taxonomy. In *Completing the Inventory of the Solar System*, ASP Conf. Ser. **107**, T.W. Rettig and J.M. Hahn, eds., pp. 173-192

Licandro, J., Tancredi, G., Lindgren, M., Rickman, H., and Hutton, R.G., 2000. CCD photometry of cometary nuclei, I: Observations from 1990–19951. *Icarus* **147**, 161 – 179.

Lisse, C.M. 2002. On the role of dust mass loss in the evolution of comets and dusty disk systems. *Earth, Moon, and Planets* **90**, 497-506.

Lisse, C.M., Fernandez, Y.R., Kundu, A., A'Hearn, M.F., Dayal, A., Deutsch, L.K., Fazio, G.G., Hora, J.L., Hoffmann, W.F., 1999. The nucleus of comet Hyakutake (C/1996 B2). *Icarus* **140**, 189-204.

Lisse, C.M., A'Hearn, M.F., Groussin, O., Fernandez, Y.R., Belton, M.J.S., van Cleeve, J.E., Charmandaris, V., Meech, K.J., and McGleam, C., 2005. Rotationally resolved 8 - 35 mm Spitzer Space Telescope observations of the nucleus of comet 9P/Tempel 1. *Astrophys. J. Lett.* **625**, L139 - L142.

Lisse, C.M.. VanCleve, J., Adams, A.C., A'Hearn, M.F., Fernández, Y.R., Farnham, T.L., Armus, L., Grillmair, C.J., Ingalls, J., Belton, M.J.S., Groussin, O., McFadden, L.A., Meech, K.J., Schultz, P.H., Clark, B.C., Feaga, L. M., and Sunshine, J.M. 2006. Spitzer Spectral Observations of the Deep Impact Ejecta, *Science* **313**, 635 – 640 and Supplementary On-line Material.

Lowry, S.C. and Fitzsimmons, A. 2001. CCD photometry of distant comets II. *Astron. Astrophys.* **365**, 204 – 213.

Reach, W.T., Kelley, M.S., and Sykes, M.V. 2007. A survey of debris trails from short-period comets. *Icarus* **191**, 298-322.

Reach, W.T., Vaubaillon, J., Kelley, M.S., Sykes, M.V., and Lisse, C.M. 2009. "Distribution and Properties of Fragments and Debris from the Split Comet 73P/Schwassmann-Wachmann 3 in 2006 as Revealed by Spitzer", *Icarus* (accepted May 2009, published on-line as arXiv:0905.3162)

Richardson, J.E., Melosh, H.J., Lisse, C.M., and Carcich, B. 2007. A ballistic analysis of the Deep Impact ejecta plume: Determining comet Tempel 1's gravity, mass, and density. *Icarus* **190**, 357-390.

Schleicher, D. G. 2007. *Deep Impact*'s target Comet 9P/Tempel 1 at multiple apparitions: Seasonal and secular variations in gas and dust production. *Icarus* **191**, 322-338.

Snodgrass, C., Lowry, S.C., and Fitzsimmons, A. 2008. Optical observations of 23 distant Jupiter Family Comets, including 36P/Whipple at multiple phase angles. *MNRAS* **385**, 737 - 756.

Snodgrass, C., Meech, K.J., and Hainaut, O.R. 2009. The nucleus of 103P/Hartley2, target of the EPOXI mission. *Astron. Astrophys.*, in prep.

Spencer, J.R., Lebofsky, L.A., and Sykes, M.V. 1989. Systematic biases in radiometric diameter determinations. *Icarus* **78**, 337-354.

Sykes, M.V. and Walker, R.W., 1992. Cometary dust trails : I. Survey. *Icarus* **95**, 180 - 210.

Thomas, P.C., *et al.* 2007. The shape, topography and geology of comet 9P/Tempel 1 from Deep Impact observations. *Icarus* **187,** 4 - 15.

Toth, I. and Lisse, C.M. 2006. On the rotational breakup of cometary nuclei and centaurs. *Icarus* **181**, 162 - 177.

Toth, I., Lamy, P.L., Weaver, H.A., Noll, K.S., and Mutchler, M.J. 2008. Hubble Space Telescope Observations of Fragment C of the Split Comet 73P/Schwassmann-Wachmann 3 in 2001 and 2006. *Bull.A.A.S.* **40**, 394

Weaver, H.A., Lisse, C.M., Mutchler, M., Lamy, P.L., Toth, I., Reach, W.T., and Vaubaillon, J., 2008. Hubble Investigation of the B and G Fragments of Comet 73P/Schwassmann-Wachmann 3. *Asteroids, Comets, and Meteorites Meeting 2008*, abstract #**8248**

Werner, M.W., *et al.* 2004. The Spitzer Space Telescope Mission. *Astrophys. J. Supp.,***154**, 1 – 9.

Yeomans, D.K., Chodas, P.W., Sitarski, G., Szutowitcz, S., and Królikowska, M. 2004. Cometary orbit determination and non-gravitational forces. In *Comets II*. Eds. M.C. Festou, H.U. Keller, and H.A. Weaver. University of Arizona Press, Tucson. pp. 137-151.

# 7. Figures

**Figure 1** – Spitzer 22 µm imagery of comet 103P/Hartley 2. Each image shown is the registered and added stack of 200 individual exposures. The compass in the lower left indicates equatorial north (N), equatorial east (E), the direction to the Sun ($\odot$), the expected direction of the comet's motion from its ephemeris (µ), and the negative of the heliocentric velocity vector (-v). The two images show the same patch of sky and the comet is identified through its apparent speed and direction of motion. Along with the comet's nucleus, the trail was detected as a linear feature lying along the PA of –v, 271º. The comet's tail, had it been present, would have been aligned closely to the projected anti-solar direction of 110º PA.

**Figure 2** – Spatial profiles of the images of comet 103P/Hartley 2 shown in Figure 1. Unfilled symbols indicate pixels within 20º azimuth of the trail direction; filled symbols show pixels everywhere else. The two profiles have been shifted vertically for clarity; note the ordinate values. Trail pixels show flux at roughly 0.036 MJy/sr above background. The uncertainty on the baseline is dominated by background fluctuations, and amounts to ±0.013 MJy/sr (1σ). For a peak signal of 0.15 MJy/sr above the background, the signal-to-noise ratio in the peak pixel is about 11.

**Figure 3 -** Histograms of active areas (top) and fractional active areas (bottom) for Jupiter-family comets, adapted from the work of A'Hearn *et al.* (1995). Unhighlighted data are from their Figure 6. The light grey datum refers to comet Tempel 1 (Lisse *et al.* 2005), and the dark grey datum refers to comet Hartley 2, as reported in this work. While comet Hartley 2 has a not-uncommon total active area, its fractional active area is quite high.

**Figure 1**

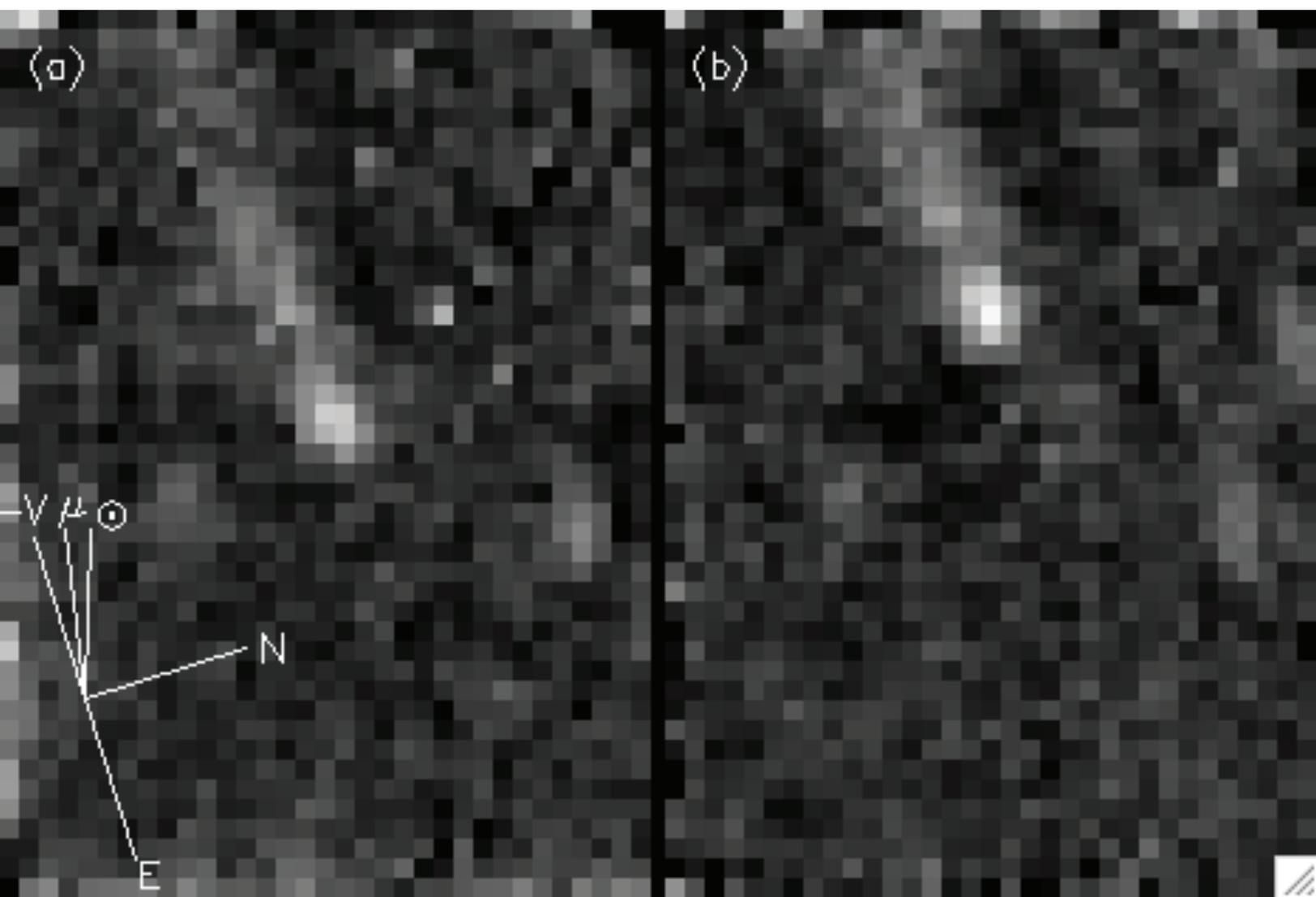

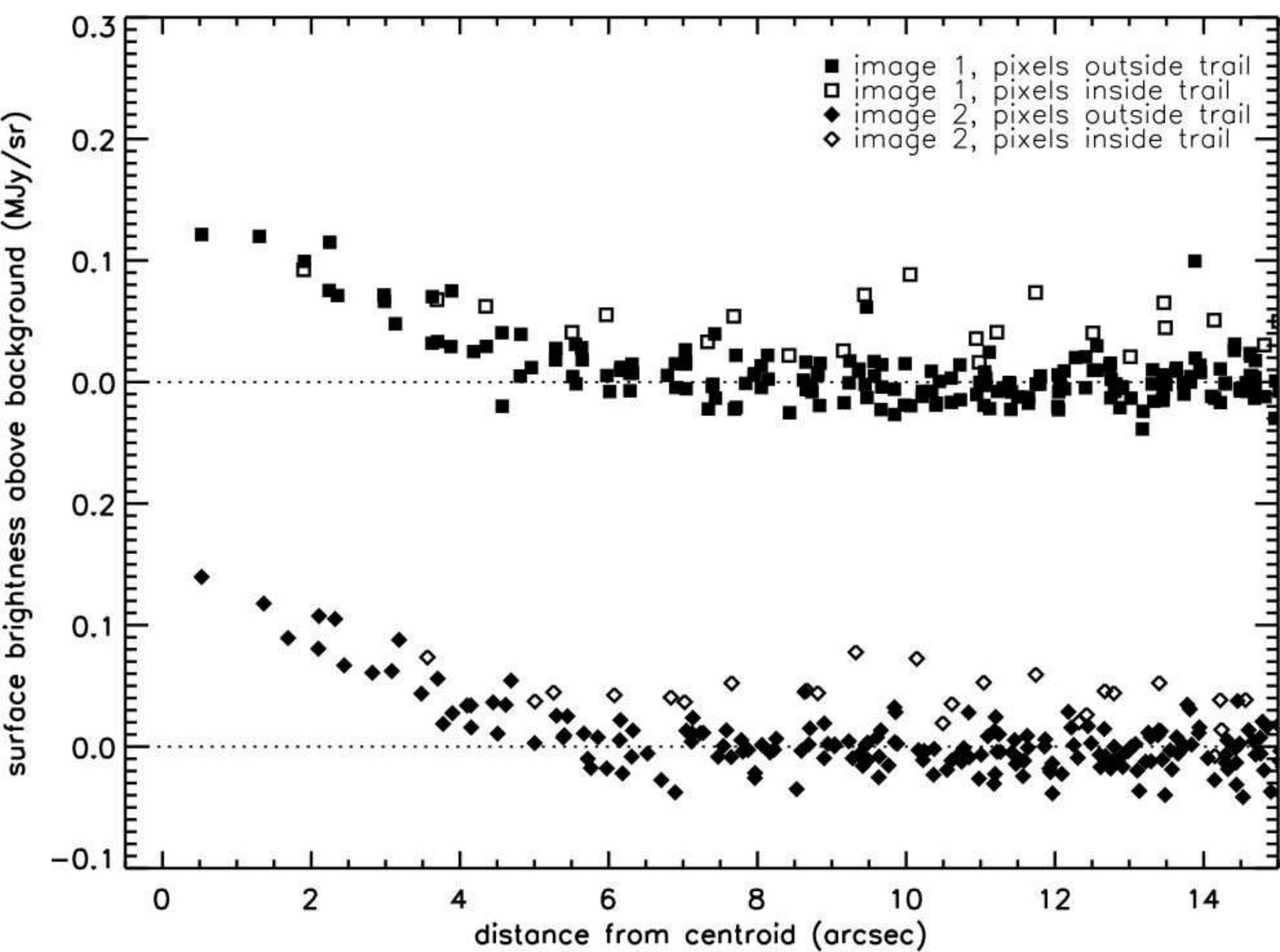

**Figure 3**

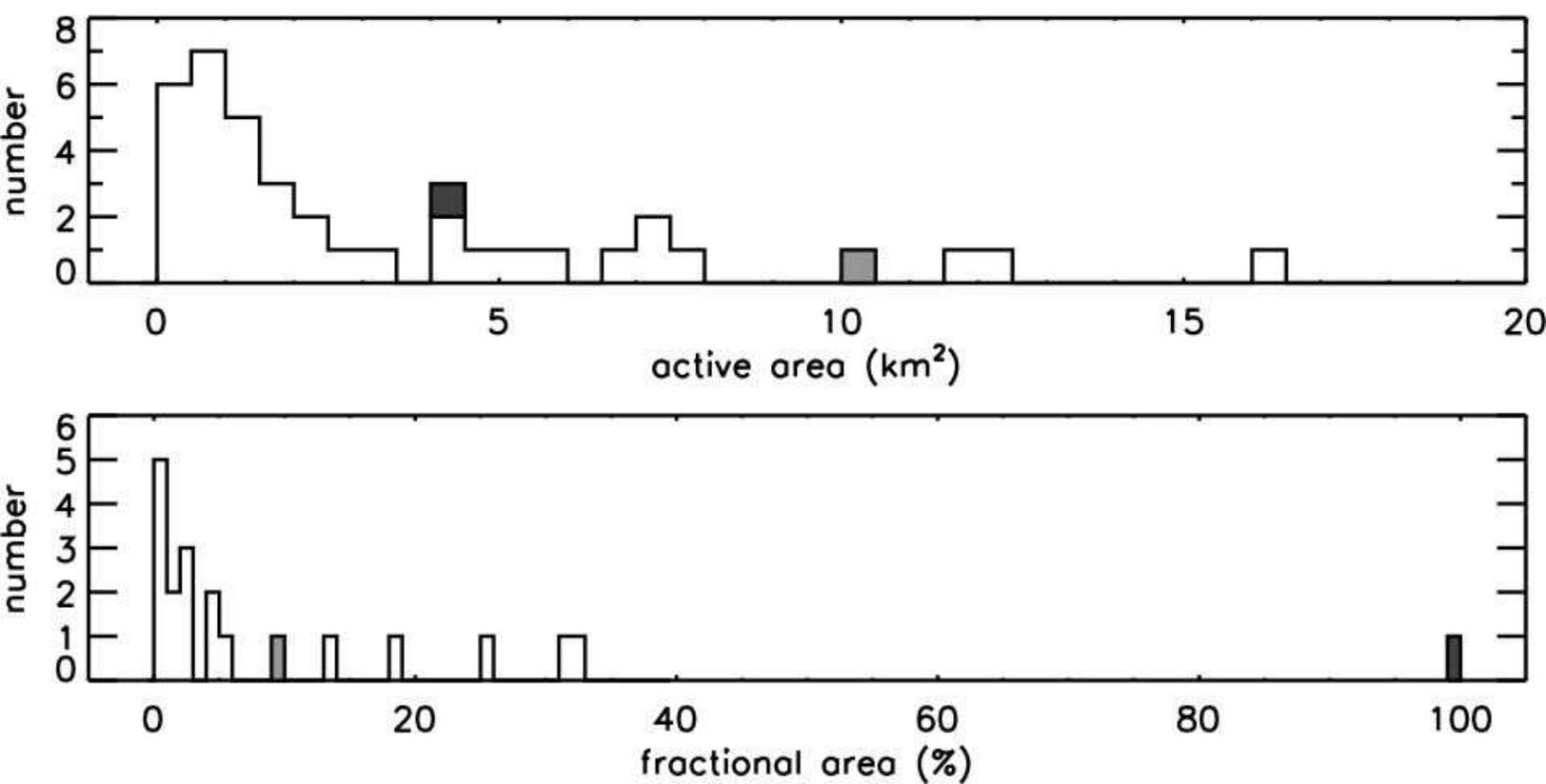